\documentclass[conference]{IEEEtran}
\IEEEoverridecommandlockouts
\usepackage{color}
\usepackage{graphicx}
\usepackage{amsmath,amssymb,amsthm}
\usepackage{multirow,bigdelim}
\usepackage{latexsym}
\usepackage{cite}

\newtheorem{theorem}{Theorem}

\newcommand{\bP}{\mathbf{P}}
\newcommand{\bW}{\mathbf{W}}

\newcommand{\bS}{\mathbf{S}}

\newcommand{\bF}{\mathbf{F}}
\usepackage{algorithm,algcompatible}
\usepackage{algorithmicx}
\usepackage{algpseudocode}
\algdef{SE}[DOWHILE]{Do}{doWhile}{\algorithmicdo}[1]{\algorithmicwhile\ #1}%
\algnewcommand\INPUT{\item[\textbf{Input:}]}%
\algnewcommand\OUTPUT{\item[\textbf{Output:}]}%
\newcommand{\bGamma}{\mathbf{\Gamma}}
\newcommand{\Vor}{\ensuremath{\mathcal{V}}}
\newcommand{\R}{\ensuremath{\mathcal{R}}}

\ifCLASSINFOpdf
   \graphicspath{{../jpg/}{../pdf/}{../jpeg/}{../eps/}}
   \DeclareGraphicsExtensions{.jpg,.pdf,.jpeg,.png,.eps}
\else
  \DeclareGraphicsExtensions{.eps}
\fi

\ifCLASSOPTIONcompsoc
  \usepackage[caption=false,font=normalsize,labelfont=sf,textfont=sf]{subfig}
\else
  \usepackage[caption=false,font=footnotesize]{subfig}
\fi
\ifCLASSOPTIONcompsoc
\else
\fi

\usepackage{setspace}

\makeatletter
\newcommand\fs@betterruled{%
  \def\@fs@cfont{\bfseries}\let\@fs@capt\floatc@ruled
  \def\@fs@pre{\vspace*{5pt}\hrule height.8pt depth0pt \kern2pt}%
  \def\@fs@post{\kern2pt\hrule\relax}%
  \def\@fs@mid{\kern2pt\hrule\kern2pt}%
  \let\@fs@iftopcapt\iftrue}
\floatstyle{betterruled}
\restylefloat{algorithm}
\makeatother

\def\BibTeX{{\rm B\kern-.05em{\sc i\kern-.025em b}\kern-.08em
    T\kern-.1667em\lower.7ex\hbox{E}\kern-.125emX}}

\begin{document}
\title{Energy Efficient Node Deployment in Wireless Ad-hoc Sensor Networks}
\author{
  \IEEEauthorblockN{Jun Guo, Saeed Karimi-Bidhendi, Hamid Jafarkhani}
  \IEEEauthorblockA{Center for Pervasive Communications \& Computing\\
                    University of California, Irvine\\
                    Email: \{guoj4, skarimib, hamidj\}@uci.edu}
}

\maketitle

\begin{abstract}
We study a wireless ad-hoc sensor network (WASN) where $N$ sensors gather data from the surrounding environment and transmit their sensed information to $M$ fusion centers (FCs) via multi-hop wireless communications. This node deployment problem is formulated as an optimization problem to make a trade-off between the sensing uncertainty and energy consumption of the network. Our primary goal is to find an optimal deployment of sensors and FCs to minimize a Lagrange combination of the sensing uncertainty and energy consumption. To support arbitrary routing protocols in WASNs, the routing-dependent necessary conditions for the optimal deployment are explored. Based on these necessary conditions, we propose a routing-aware Lloyd algorithm to optimize node deployment. Simulation results show that, on average, the proposed algorithm outperforms the existing deployment algorithms.
\end{abstract}


\section{Introduction}\label{sec:intro}
Recent developments in wireless communications, digital electronics and computational power have enabled a large number of applications of wireless ad-hoc sensor networks (WASNs) in various fields such as agriculture, industry and military. In a general WASN, spatially dispersed and dedicated sensors collect data, e.g. temperature, sound, pressure and radio signals from the physical environment, and then forward the gathered information to one or more fusion centers (FCs) via wireless communications.

In order to collect accurate data from the physical surroundings, high sensing quality or sensitivity is required. In general, sensing quality diminishes as the distance between the sensor and target point increases \cite{SFMM,SF,AG,AB,TPL,RS}. Thus, two distance-dependent measures, i.e., sensing coverage \cite{SFMM,AJ,YK,YY,ICC} and sensing uncertainty \cite{AB,Erdem1,GJ,GJICC,YBZ,JS,FJSY} are widely studied in the literature to evaluate the sensing quality. In the binary coverage model \cite{SFMM,AJ,YK,YY,ICC}, each sensor node can only detect the events within the sensing radius $R_s$. Then, sensing coverage represents the percentage of events that is covered by at least one sensor \cite{SFMM,AJ,YK,YY}. But when the number of sensors is large enough, such that the full coverage can be achieved, coverage degree \cite{ICC}, i.e., the minimum number of sensors that any event can be detected by, is a better sensing quality measure. Another common model, centroidal Voronoi tessellation, formulates the sensing quality as a source coding problem with sensing uncertainty as its distortion \cite{AB,Erdem1,GJ,GJICC,YBZ,JS,FJSY}.

Energy efficiency is another key metric in WASNs as it is inconvenient or even infeasible to recharge the batteries of numerous and densely deployed sensors. In general, wireless communication, sensing and data processing are three primary energy consumption components of a node. However, in many WASN applications, wireless communication among nodes is more power hungry compared to other components \cite{GMMA,MS}. Therefore, wireless communication dominates the node energy consumption in practice.

There are four primary energy saving methods for WASNs in the literature: (1) topology control \cite{XYY,XW}, in which unnecessary energy consumption is reduced by properly switching the nodes' states (sleeping or working); (2) clustering \cite{OYSF,VP} which is used to balance the energy consumption among nodes in one-hop communication models by iteratively selecting cluster heads; (3) energy-efficient routing \cite{JL,Ford,Bellman}, a widely used method that attempts to find the optimal routing paths to forward data to FCs while the communication cost between two nodes are held fixed; and (4) deployment optimization that plays an important role in the energy consumption of WASNs since the communication cost between two nodes depends on their distance. Our previous works \cite{JGTCOM,SJH} proposed Lloyd-like algorithms to save communication energy in homogeneous and heterogeneous WASNs by optimizing the node deployment. Nonetheless, a pre-existing network infrastructure, which only includes two-hop communications, is a basic assumption in \cite{JGTCOM,SJH}. Compared to one-hop and two-hop communications, the generalized multi-hop communications can, on average, reduce the transmission distance and save more energy. However, to the best of our knowledge, the optimal node deployment with generalized multi-hop communications in WASNs is still an open problem.

In this paper, we study the node deployment problem in WASNs with arbitrary multi-hop routing algorithms. Our primary goal is to find the optimal FC and sensor deployment to minimize both sensing uncertainty and total energy consumption of the network. By deriving the routing-dependent necessary conditions of the optimal deployments in such WASNs, we design a Lloyd-like algorithm to deploy nodes.

The rest of this paper is organized as follows: In Section \ref{sec:model}, we introduce the system model and problem formulation. In Section \ref{sec:opt}, we study the optimal FC and sensor deployment for a given routing algorithm. A numerical algorithm is proposed in Section \ref{sec:algorithm} to optimize the node deployment. Section \ref{sec:simulation} presents the experimental results and Section \ref{sec:conclusion} concludes the paper.

\section{System model and problem formulation}\label{sec:model}

We consider a  wireless ad-hoc sensor network consisting of $M$ FCs and $N$ sensors over a target region $\Omega\in\mathbb{R}^2$. For convenience, we define $\mathcal{I}_{S}=\{1,\dots,N\}$ to be the set of node indices for sensors and $\mathcal{I}_{F}=\{N+1,\dots,N+M\}$ to be the set of node indices for FCs.
When $i\in\mathcal{I}_{S}$, Node $i$ refers to Sensor $i$; however, when $i\in\mathcal{I}_{F}$, Node $i$ refers to FC $(i-N)$. Let $\bP=(p_1,\dots,p_N,p_{N+1},\dots,p_{N+M})^{T}\in\mathbb{R}^{(N+M)\times2}$ be the node deployment, where $p_i\in\Omega$ is Node $i$'s location. Throughout this paper, we assume that each event is sensed by only one sensor. Therefore, for each node deployment $\mathbf{P}$, there exists a cell partitioning $\bW=(W_1,\dots,W_{N})^{T}$ comprised of $N$ disjoint subsets of $\mathbb{R}^2$ whose union is $\Omega$, and each sensor, say $i$, only monitors the events occurred in the cell $W_i\subseteq \Omega$. The spatial density function that reflects the frequency of random events taking place over the target region $\Omega$ is denoted via a continuous and differentiable function $f(\omega):\Omega\to\mathbb{R}^{+}$. Let $\bGamma(\bW)=\left(\Gamma_1(\bW),\dots,\Gamma_N(\bW)\right)^{T}\in\mathbb{R}^{N}$ where $\Gamma_i(\bW)$ is the amount of data generated at Node (Sensor) $i$ in the unit time. Since Sensor $i$ detects the events in the region $W_i$, $\Gamma_i(\bW)$ is proportional to the volume of $W_i$, i.e.,  $\Gamma_i(\bW)=\kappa\int_{W_i}f(\omega)d\omega$ where $\kappa$ is a constant, \cite{JGTCOM}.

According to \cite{JL}, this WASN can be modeled as a directed acyclic graph $\mathcal{G}(\mathcal{I}_{S}\bigcup \mathcal{I}_{F},\mathcal{E})$ where $\mathcal{E}$ is the set of directed links $(n,k)$ such that $n\in\mathcal{I}_{S}$ and $k\in\mathcal{I}_{S}\bigcup\mathcal{I}_{F}$.
In particular, sensors and FCs are source nodes and sink nodes of this network, respectively, and there is no cycle in the flow network since each cycle can be eliminated by reducing the flows along the cycle without influencing the in-flow and out-flow links to that cycle. We define $\bF=[F_{i,j}]_{N\times (N+M)}$ to be the flow matrix, where $F_{i,j}$ is the amount of data transmitted through the link $(i,j)$ in the unit time. Since $\bF$ depends on the cell partitioning $\bW$, we can define the normalized flow matrix as follows:
\begin{equation}
\bS=
{\begin{array}{rccccll}
 & \multicolumn{4}{l}{\overbrace{\hspace{12em}}^{N+M}} & \\
\ldelim[{5}{5pt}[]\!\!\! & \textcolor{black}{s_{1,1}} & \textcolor{black}{s_{1,2}} &
\cdots & \textcolor{black}{s_{1,N+M}} &  \!\!\!\!\rdelim]{5}{4pt}[] & \!\!\!\!\rdelim\}{5}{5pt}[\emph{N},]\\
 & \textcolor{black}{s_{2,1}} & \textcolor{black}{s_{2,2}} &
 \cdots & \textcolor{black}{s_{2,N+M}} & \\ 
 & \vdots & \vdots & \ddots & \vdots & \\
 & \textcolor{black}{s_{N,1}} & \textcolor{black}{s_{N,2}} &
 \cdots & \textcolor{black}{s_{N,N+M}} & \\
\end{array}}
\end{equation}

\noindent where $s_{i,j}\triangleq\frac{F_{i,j}}{\sum_{j=1}^{N+M}F_{i,j}}$ is the data ratio that Node $i$ transmits to Node $j$. The normalized flow matrix $\bS$ satisfies the following properties:

\noindent(a) $s_{i,j}\in[0,1]$;\footnote{For time-invariant routing algorithms, such as Bellman-Ford Algorithm \cite{Ford,Bellman}, the flows construct a tree-structured graph in which each node has only one successor.
Under such a circumstance, the normalized flow from Node $i$ to Node $j$ is either $0$ or $1$, i.e., $s_{i,j}\in\{0,1\}$.
However, the time-variant routing algorithms, such as Flow Augmentation Algorithm \cite{JL}, will generate different flows during different time periods. As a result, the overall normalized flow from Node $i$ to Node $j$ can be a real number between $0$ and $1$, i.e., $s_{i,j}\in[0,1]$.}\\
(b) $\sum_{j=1}^{N+M}s_{i,j}=1, \forall i\in\{1,\dots,N\}$;\\
(c) No cycle: if there exists a path $l_0\!\to\!l_1\!\to\!\cdots\!\to\!l_K$, i.e., $\prod_{k=1}^{K}s_{l_{k-1},l_k}>0$, then we have $s_{l_K,l_0}=0$. In particular, $s_{ii}=0, \forall i\in\{1,\dots,N\}$.

Since the flow $F_{i,j}$ can be determined by the cell partitioning $\bW$ and normalized flow matrix $\bS$, in the remaining of this paper we use $\bF(\bW, \bS)$ instead of $\bF$. Let $F_i(\bW, \bS)\triangleq\sum_{j=1}^{N+M}F_{i,j}(\bW,\bS)$ be the total flow originated from Node $i$. Since the in-flow to each sensor, say $i$, should be equal to the out-flow, we have $\sum_{j=1}^{N}F_{j,i}(\bW,\bS)+\Gamma_i(\bW)=\sum_{j=1}^{N+M}F_{i,j}(\bW,\bS).$ In what follows, we provide an example to elucidate how to calculate $F(\bW,\bS)$ in terms of $\bW$ and $\bS$.

\begin{figure}[!htb]
\setlength\abovecaptionskip{0pt}
\setlength\belowcaptionskip{0pt}
\centering
\includegraphics[width=3in]{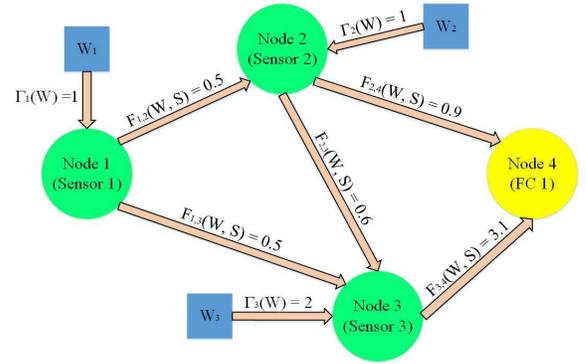}
\captionsetup{justification=justified}
\caption{\small{Example 1}}
\label{Example1}
\end{figure}

\emph{Example 1.} We consider a WASN with three sensor nodes and one FC, i.e., $N=3$ and $M=1$.
The parameter $\kappa$ is set to $4$.
For a cell partitioning $\bW$ with cell volumes $v_1(W)=v_2(W)=0.25$, $v_3(W)=0.5$, and the normalized flow matrix
    $\bS =
    \begin{array}{rccccl}
       \ldelim[{3}{5pt}[]
       & 0 & 0.5 & 0.5 & 0 &
       \rdelim]{3}{5pt}[]\\
       & 0 & 0 & 0.4 & 0.6 & \\
       & 0 & 0 & 0 & 1 & \\
    \end{array}$,
the corresponding flow network is illustrated in Fig. \ref{Example1}.

The amount of data generated from each sensor node can be calculated as: $\Gamma_1(\bW)\!=\!\kappa v_1(\bW)\!=\!1$, $\Gamma_2(\bW)\!=\!\kappa v_2(\bW)\!=\!1$, and $\Gamma_3(\bW)\!=\!\kappa v_3(\bW)\!=\!2$. As a leaf node, Sensor 1 does not receive data from any other sensor nodes, and only transmits its sensed data; thus, $F_1(\bW,\bS)\!=\!\Gamma_1(\bW)\!=\!1$. The flows from Sensor $1$ are then $F_{1,2}(\bW,\bS)\!=\!s_{1,2}\!\times\!F_{1}(\bW,\bS)\!=\!0.5$ and $F_{1,3}(\bW,\bS)\!=\!s_{1,3}\!\times\!F_{1}(\bW,\bS)\!=\!0.5$, respectively. Sensor $2$'s flows come from $F_{1,2}(\bW,\bS)$ and the data gathered from the region $W_2$. Hence, $F_2(\bW,\bS)\!=\!\Gamma_2(\bW)\!+\!F_{1,2}(\bW,\bS)\!=\!1.5$. Therefore, the flows from Sensor $2$ are $F_{2,3}(\bW,\bS)\!=\!s_{2,3}\!\times\!F_{2}(\bW,\bS)\!=\!0.6$ and $F_{2,4}(\bW,\bS)\!=\!s_{2,4}\!\times\!F_{2}(\bW,\bS)\!=\!0.9$. Similarly, for Sensor $3$, we have $F_3(\bW,\bS)\!=\!\Gamma_3(\bW)\!+\!F_{1,3}(\bW,\bS)\!+\!F_{2,3}(\bW,\bS)\!=\!3.1$; hence, the unique flow from Sensor $3$ is $F_{3,4}(\bW,\bS)\!=\!s_{3,4}\times F_3(\bW,\bS)\!=\!3.1$.

Now, we have enough materials to formulate the energy consumption in a WASN. In general, FCs are equipped with reliable energy sources and their energy consumption is not the main concern. In what follows, we focus on the sensor energy consumption. The average power consumption (Watts) through link $(i,j)$ consists of two components: \emph{average-transmitter-power}, $\overline{\mathcal{P}}^T_{i,j}$, and \emph{average-receiver-power}, $\overline{\mathcal{P}}^R_{i,j}$. Sensor $i$'s \emph{average-transmitter-power} through link $(i,j)$ can be expressed as $\overline{\mathcal{P}}^T_{i,j}=\mathcal{P}^T_{i,j}\times r_{i,j}$, where  $\mathcal{P}^T_{i,j}$ is Sensor $i$'s \emph{instant-transmitter-power} through link $(i,j)$, and $r_{i,j}$ is the \emph{link-busy-ratio}, i.e., the percentage of time that data is going through link $(i,j)$.
To achieve a reliable communication between two nodes, the \emph{instant-transmission-power} (Joules/second) required for Node $i$ to transmit data to Node $j$ should be set to $\mathcal{P}^T_{i,j}=\xi_{i,j}||p_i-p_j||^{\alpha}$, where $||.||$ denotes the Euclidean distance, $\alpha$ is the path-loss parameter, $\xi_{i,j}$ is a constant determined by Node $i$'s antenna gain, Node $j$'s antenna gain, and Node $j$'s SNR threshold. In this paper, we consider free-space path-loss, i.e., $\alpha=2$ and homogeneous sensors and FCs, i.e., $\xi_{i,j}=\xi$. Therefore, the \emph{instant-transmitter-power} through link $(i,j)$ is modeled as $\mathcal{P}^T_{i,j}=\xi||p_i-p_j||^{2}$.
It is reasonable to assume that the link $(i,j)$
has idle time, and it is busy only when there is data to transmit from Node $i$ to Node $j$.
Hence, link $(i,j)$'s \emph{link-busy-ratio} can be
written as $\frac{F_{i,j}(\bW,\bS)\times T/\zeta_{i,j}}{T}=\frac{F_{i,j}(\bW,\bS)}{\zeta_{i,j}}$, where $\zeta_{i,j}$ is the \emph{instant-data-rate} through link $(i,j)$ which is determined by the bandwidth of $(i,j)$. In this paper, we assume that all links have the same bandwidth, i.e., $\zeta_{i,j}=\zeta$.
Therefore, Sensor $i$'s \emph{average-transmitter-power} through link $(i,j)$ can be rewritten as
$\overline{\mathcal{P}}^T_{i,j}=\beta||p_i-p_j||^{2}F_{i,j}(\bW,\bS)$, where $\beta=\frac{\xi}{\zeta}$. According to \cite{YYHF}, Sensor $j$'s
\emph{average-receiver-power} through link $(i,j)$ can be modeled as $\overline{\mathcal{P}}^{R}_{i,j}=\rho F_{i,j}(\bW,\bS)$, where $\rho$ is a constant coefficient for receiving data.
In sum, the average power consumption over link $(i,j)$ can be written as
\vspace{-2pt}
\begin{equation}
\begin{aligned}
&\overline{\mathcal{P}}_{i,j}(\bP,\bW,\bS)=\overline{\mathcal{P}}^{T}_{i,j}+\overline{\mathcal{P}}^{R}_{i,j}\\
&=\begin{cases}
\left(\beta||p_i-p_j||^{2}+\rho\right)F_{i,j}(\bW,\bS), &\quad j\in\mathcal{I}_{S}\\
\left(\beta||p_i-p_j||^{2}\right)F_{i,j}(\bW,\bS),  &\quad j\in\mathcal{I}_{F}\\
\end{cases}
\end{aligned}
\end{equation}
and the total power consumption can be written as
\vspace{-2pt}
\begin{equation}
\begin{aligned}
    &\overline{\mathcal{P}}(\bP,\bW,\bS)
    =\sum_{i=1}^{N}\sum_{j=1}^{N+M}\overline{\mathcal{P}}_{i,j}(\bP,\bW,\bS)\\
    &=\sum_{i=1}^{N}\!\left[\!\sum_{j=1}^{N+M}\!\!\!\beta||p_i\!-\!p_j||^{\!2}F_{i,j}(\bW,\bS)\!+\!\rho\!\sum_{j=1}^{N}\!\!F_{i,j}(\bW,\bS)\right].
    \label{powerConsumption1}
\end{aligned}
\end{equation}
According to \cite{AB,Erdem1,GJ,GJICC,YBZ,JS,FJSY}, for a given node deployment $\bP$ and cell partitioning $\bW$, the sensing uncertainty can be formulated as:
\vspace{-5pt}
\begin{equation}
\mathcal{H}(\bP,\bW)=\sum_{n=1}^{N}\int_{W_n(\bP)}\|p_n-\omega\|^2f(\omega)d\omega.
\label{sensingUncertainty}
\end{equation}
Taking sensing uncertainty and energy consumption into consideration,
the objective (cost) function is then defined as the Lagrangian function of Eqs. (\ref{powerConsumption1}) and (\ref{sensingUncertainty}):
\begin{equation}
    \begin{aligned}
    &D(\bP,\bW,\bS)=\mathcal{H}(\bP,\bW)+\lambda \overline{\mathcal{P}}(\bP,\bW,\bS)\\
    &=\sum_{i=1}^{N}\int_{W_i}\|p_i-\omega\|^2f(\omega)d\omega+\lambda\rho\sum_{i=1}^{N}\sum_{j=1}^{N}F_{i,j}(\bW,\bS)\\
    &+ \sum_{i=1}^{N}\sum_{j=1}^{N+M}\left(\lambda\beta||p_i\!-\!p_j||^{2}\right)F_{i,j}(\bW,\bS),
    \end{aligned}
    \label{distortion1}
\end{equation}
where $\lambda\ge0$ is the Lagrangian multiplier.
Our main goal in this paper is to minimize the cost function defined in (\ref{distortion1}) over the node deployment $\bP$, cell partitioning $\bW$, and the normalized flow matrix $\bS$.

\section{Optimal node deployment in WASNs}\label{sec:opt}

In this section, we study the optimality conditions for node deployment $\bP$, cell partitioning $\bW$ and the normalized flow matrix $\bS$ to minimize the objective function defined in (\ref{distortion1}). Let
\begin{equation}
    \begin{aligned}
    \!e_{i,j}(\bP)\triangleq\frac{\overline{\mathcal{P}}_{i,j}(\bP,\bW,\bS)}{F_{i,j}(\bW,\bS)}
    \!=\!\begin{cases}
\beta||p_i\!-\!p_j||^{2}\!+\!\rho, & \!j\in\mathcal{I}_{S}\\
\beta||p_i\!-\!p_j||^{2},  &\!j\in\mathcal{I}_{F}\\
\end{cases}
    \end{aligned}
\end{equation}
be the Link $(i,j)$'s energy cost (Joules/bit). Without loss of generality, we assume that Sensor $i$'s sensing data goes through $K_i$ paths $\{L^{(i)}_{k}(\bS)\}_{k\in\{1,\dots,K_i\}}$, where $L^{(i)}_{k}(\bS)=l^{(i)}_{k,0}\to l^{(i)}_{k,1}\to \cdots\to l^{(i)}_{k,J^{(i)}_k}$, $l^{(i)}_{k,0}=i$, $l^{(i)}_{k,J^{(i)}_k}\in\mathcal{I}_{F}$, and $J^{(i)}_k$ is the number of nodes on the $k$-th path except Node $i$.
Then, the data rate (bit/s) and the path cost (Joules/bit) corresponding to the $k$-th path can be written as
\begin{equation}\label{data_rate}
    \mu_k^{(i)}(\bW,\bS)=F_i(\bW,\bS)\prod_{j=1}^{J^{(i)}_k}s_{l^{(i)}_{k,j-1},l^{(i)}_{k,j}}
\end{equation}
and
\begin{equation}\label{path_loss}
    \overline{e}^{(i)}_k(\bP,\bS)=\sum_{j=1}^{J^{(i)}_k}e_{l^{(i)}_{k,j-1},l^{(i)}_{k,j}}(\bP)
\end{equation}
respectively. Note that $\sum_{k}\mu_k^{(i)}(\bW, \bS) = F_i(\bW,\bS)$ which means the data from Node $i$ eventually reach a FC. Sensor $i$'s \emph{power coefficient}, denoted as $g_i(\bP, \bS)$, is then defined to be the energy consumption (Joules/bit) for transmitting $1$ bit data from Sensor $i$ to the FCs, i.e, we have\footnote{The term $F_i(\bW,\bS)$ is canceled in Eq. (\ref{g}), indicating that power coefficient $g_i(\bP,\bS)$ is independent of $\bW$.}:
\begin{equation}
\begin{aligned}
    &g_i(\bP,\bS)=\frac{\sum_{k=1}^{K_i}\mu_k^{(i)}(\bW,\bS)\overline{e}^{(i)}_k(\bP,\bS)}{F_i(\bW,\bS)}\\
    \!\!=&\!\sum_{k=1}^{K_i}\!\!\left[\!\prod_{j=1}^{J^{(i)}_k}\!s_{l^{(i)}_{k,j-1},l^{(i)}_{k,j}}\!\!\!\left(\sum_{j=1}^{J^{(i)}_k}\!\beta\!\left\|p_{l^{(i)}_{k,j-1}}\!\!\!\!-p_{l^{(i)}_{k,j}}\!\right\|^{2}\!\!\!+\!\rho\!\left(\!J^{(i)}_k\!\!\!-\!\!1\!\right)\!\right)\!\right].
\end{aligned}
\label{g}
\end{equation}
In what follows, we provide an example to clarify how to calculate the sensor power coefficients.

\emph{Example 2.} Consider the WASN described in Fig. \ref{Example1}, and let $\bP=((0,0), (0,1), (1,0), (1,1))$, $\beta\!=\!1$ and $\rho\!=\!1$. We aim to find Sensor 1's power coefficient $g_1(\bP,\bS)$. The link energy costs for this network can be calculated as $e_{1,2}(\bP)=e_{1,3}(\bP)\!=\!2$, $e_{2,3}(\bP)\!=\!3$, and $e_{2,4}(\bP)\!=\!e_{3,4}(\bP)\!=\!1$. Note that Sensor $1$'s data goes through the following 3 paths: $L^{(1)}_{1}(\bS)\!=\!1\!\to\!2\!\to\!4$, $L^{(1)}_{2}(\bS)\!=\!1\!\to\!3\!\to\!4$, and
$L^{(1)}_{3}(\bS)\!=\!1\!\to\!2\!\to\!3\!\to\!4$. The data rate through the above paths are, respectively,
$\mu_1^{(1)}(\bW,\bS)\!=\!F_1(\bW,\bS)\!\times\!s_{1,2}\!\times\!s_{2,4}\!=\!0.3F_1(\bW,\bS)$,
$\mu_2^{(1)}(\bW,\bS)\!=\!F_1(\bW,\bS)\!\times\!s_{1,3}\!\times\!s_{3,4}\!=\!0.5F_1(\bW,\bS)$, and
$\mu_3^{(1)}(\bW,\bS)\!=\!F_1(\bW,\bS)\!\times\!s_{1,2}\!\times\!s_{2,3}\!\times\!s_{3,4}\!=\!0.2F_1(\bW,\bS)$.
Moreover, we can calculate the path costs using Eq. (\ref{path_loss}) as follows:
$\overline{e}^{(1)}_{1}(\bP)=e_{1,2}(\bP)+e_{2,4}(\bP)=3$, $\overline{e}^{(1)}_{2}(\bP)=e_{1,3}(\bP)+e_{3,4}(\bP)=3$, and
$\overline{e}^{(1)}_{3}(\bP)=e_{1,2}(\bP)+e_{2,3}(\bP)+e_{3,4}(\bP)=6$.
Then, Sensor $1$'s power coefficient is $g_1(\bP,\bS)=0.3\times3+0.5\times3+0.2\times6=3.6$.

Note that the average power consumption for transmitting Sensor $i$'s data is $g_{i}(\bP,\bS)\Gamma_i(\bW)\!=\!g_{i}(\bP,\bS)\kappa\int_{\bW_i}\!\!f(\omega)d\omega$.
Thus, the total power consumption (\ref{powerConsumption1}) can be rewritten as:
\begin{align}
    \overline{\mathcal{P}}(\bP,\bW,\bS)
    =\sum_{i=1}^{N}g_{i}(\bP,\bS)\kappa\int_{\bW_i}\!\!f(\omega)d\omega.
    \label{powerConsumption2}
\end{align}
Therefore, the cost function in (\ref{distortion1}) can be rewritten as:
\begin{equation}
    \begin{aligned}
    &D(\bP,\bW,\bS)=\mathcal{H}(\bP,\bW)+\lambda \overline{\mathcal{P}}(\bP,\bW,\bS)\\
    &=\sum_{i=1}^{N}\int_{W_i}\left(\|p_i-\omega\|^2+\lambda\kappa g_i(\bP,\bS)\right)f(\omega)d\omega.
    \end{aligned}
    \label{distortion2}
\end{equation}

Given the node deployment $\bP$ and normalized flow matrix $\bS$, the optimal cell partitioning, also referred to as power diagrams in \cite{VD}, is equal to:
\begin{equation}
\begin{aligned}
    \Vor_i(\bP,\bS)=&\{\omega|\|p_i-\omega\|^2+\lambda\kappa g_i(\bP,\bS)\le\\
    &\|p_j-\omega\|^2+\lambda\kappa g_j(\bP,\bS), \forall j\ne i\}, i\in\mathcal{I}_{S}.
\end{aligned}
\end{equation}
Moreover, given the link costs $\{e_{ij}(\bP)\}$s and generated sensing data rates $\{\Gamma_i(\bW)\}$s, the total power consumption can be minimized by Bellman-Ford Algorithm \cite{Ford,Bellman}. For convenience, we represent the functionality of Bellman-Ford Algorithm by $\R(\bP,\bW)$, where $\bP$ and $\bW$ are inputs and $\bS$ is the output, i.e., $\R(\bP,\bW)=\arg\min_{\bS}\overline{\mathcal{P}}(\bP,\bW,\bS)$. Since sensing uncertainty $\mathcal{H}(\bP,\bW)$ is independent of $\bS$, we have:
\begin{equation}
\begin{aligned}
    \R(\bP,\bW)&=\arg\min_{\bS}\mathcal{H}(\bP,\bW)+\beta\overline{\mathcal{P}}(\bP,\bW,\bS)\\
    &=\arg\min_{\bS}D(\bP,\bW,\bS)
\end{aligned}
\end{equation}
The optimal flow matrix for the given $\bP$ and $\bW$ is then $\bF(\bW,\R(\bP,\bW))$. The following theorem provides the necessary conditions for the optimal deployment.

\begin{theorem}
The necessary conditions for the optimal deployments in the WASNs with the cost defined by Eq. (\ref{distortion1}) are
\begin{align}
&\!\!p^*_i\!=\!\frac{c^*_iv^*_i\!+\!\lambda\beta\!\sum_{j=1}^{N\!+\!M}\!F^*_{i,j}p^*_j\!+\!\lambda\beta\sum_{j=1}^{N}\!F^*_{j,i}p^*_j}{v^*_i+\lambda\beta\left(\sum_{j=1}^{N+M}F_{i,j}^*+\sum_{j=1}^{N}F^*_{j,i}\right)}, \forall i\in\mathcal{I}_{S}
\label{optP} \\
&\quad\quad\quad\quad p^*_i=\frac{\sum_{j=1}^{N}F^*_{j,i}p^*_j}{\sum_{j=1}^{N}F^*_{j,i}}, \forall i\in\mathcal{I}_{F}
\label{optQ} \\
&\quad\quad\quad\quad\quad\quad\bW^*=\Vor(\bP^*,\bS^*),
\label{optW}\\
&\quad\quad\quad\quad\quad\quad\bS^*=\R(\bP^*,\bW^*),
\label{optS}
\end{align}
where $p_i^*$ is the optimal location for Node $i$,
$\bW^*$ is the optimal cell partitioning, $\bS^*$ is the optimal normalized flow matrix,
$v^*_i(\bP^*,\bS^*)=\int_{\Vor_i(\bP^*,\bS^*)}f(\omega)d\omega$ is the Lebesgue measure (volume) of $\Vor_i(\bP^*,\bS^*)$, $c^*_i=\frac{\int_{\Vor_i(\bP^*,\bS^*)}\omega f(\omega)d\omega}{v^*_i(\bP^*,\bS^*)}$ is the geometric centroid of $\Vor_i(\bP^*,\bS^*)$, and $F^*_{i,j}=F_{i,j}(\bW^*,\bS^*)$ is the optimal flow from Node $i$ to Node $j$.
\label{theorem1}
\end{theorem}
The proof of Theorem 1 is provided in Appendix \ref{Appendix:theorem1}.
Let $\mathcal{N}^P_i(\bS)\triangleq\{j|F_{j,i}(\bW,\bS)>0\}$ be the set of Node $i$'s predecessors, and $\mathcal{N}^S_i(\bS)\triangleq\{j|F_{i,j}(\bW,\bS)>0\}$ be the set of Node $i$'s successors.
Hence, Eqs. (\ref{optP}) and (\ref{optQ}) can be simplified as
\begin{equation}
    \!\!p^*_i\!=\!\frac{c^*_iv^*_i\!+\!\lambda\beta\sum_{j\in\mathcal{N}^S_i(\bS^*)}\!F^*_{i,j}p^*_j\!+\!\lambda\beta\sum_{j\in\mathcal{N}^P_i(\bS^*)}\!F^*_{j,i}p^*_j}{v^*_i+\lambda\beta\left(\sum_{j\in\mathcal{N}^S_i(\bS^*)}F_{i,j}^*+\sum_{j\in\mathcal{N}^P_i(\bS^*)}F^*_{j,i}\right)}
\end{equation}
for each $i\in\mathcal{I}_{S}$, and
\begin{equation}
    p^*_i=\frac{\sum_{j\in\mathcal{N}^P_i(\bS^*)}F^*_{j,i}p^*_j}{\sum_{j\in\mathcal{N}^P_i(\bS^*)}F^*_{j,i}}
\end{equation}
for each $i\in\mathcal{I}_{F}$, respectively. In other words, Sensor $i$'s optimal location is a linear combination of its geometric centroid, predecessors, and successors while FC $j$'s optimal location is a linear combination of its predecessors.

\section{Routing-aware Lloyd Algorithm}\label{sec:algorithm}

Before introducing our new Routing-aware Lloyd (RL) Algorithm to solve the deployment problem, we quickly review Lloyd Algorithm \cite{Lloyd}. Lloyd Algorithm iterates between  two steps: (i) Voronoi partitioning and (ii) Moving each node to the geometric centroid of its corresponding Voronoi region. Based on Lloyd Algorithm, we use the necessary conditions in Theorem \ref{theorem1} to design RL Algorithm and optimize the node deployment in WASNs. Starting with a random initialization for node deployment $\bP$ in the target region $\Omega$, first, we design a quantizer with $N$ ($M$) points for the sensor density function and place the sensors (FCs) on the corresponding centroids to encourage an even distribution of sensors among FCs and account for a possibly poor initial node deployment. RL Algorithm then iterates between three steps: (i) Update node locations according to Eqs. (\ref{optP}) and (\ref{optQ});
(ii) Run Bellman-Ford Algorithm to obtain the flow matrix $\bF(\bP,\bW)$ and update the normalized flow matrix $\bS$ and sensor power coefficients $g_i(\bP,\bS)$; and (iii) Calculate the cell partitioning according to Eq. (\ref{optW}) and update the value of volumes $v_n$ and centroids $c_n$. The algorithm continues until the stop criterion is satisfied. The details of RL Algorithm is provided in Algorithm \ref{RELA}.

\begin{algorithm}[tb]
\smallskip
\caption{ Routing-aware Lloyd Algorithm}
\label{RELA}
\begin{algorithmic}[1]
\INPUT
Target area $\Omega$;
density function $f(\cdot)$;
initial node deployment $\bP^0$;
Lagrange multiplier $\lambda$;
stop threshold $\epsilon$.
\OUTPUT
Node deployment $\bP$;
cell partition $\bW$;
normalized flow matrix $\bS$;
cost function $D(\bP, \bW, \bS)$.
\State Run Lloyd Algorithm for Sensors and update $\{p_i\}_{i\in\mathcal{I}_{S}}$
\State Run Lloyd Algorithm for FCs and update $\{p_i\}_{i\in\mathcal{I}_{F}}$
\State Initialize the normalized flow matrix $\bS=[\mathbf{I}_{N\times N}|\mathbf{0}_{N\times M}]$
\State Calculate the power diagrams $\Vor_{i}(\bP,\bS)$, $\forall i\in\mathcal{I}_{S}$
\State Calculate the flow matrix $\bF(\bW,\bS)$
\Do
    \State Calculate the old cost function $D_{old}=D(\bP, \bW, \bS)$
    \State Calculate centroid $c_i$ and volume $v_i$ of $W_i$, $\forall i\in\mathcal{I}_{S}$
    \For{$i$ = 1 to $N$}
    \State Move Sensor $i$ to  $\frac{c_iv_i\!+\!\lambda\beta\!\sum_{j=1}^{N\!+\!M}\!F_{i,j}p_j\!+\!\lambda\beta\!\sum_{j=1}^{N}\!F_{j,i}p_j}{v_i+\lambda\beta\left(\sum_{j=1}^{N+M}F_{i,j}+\sum_{j=1}^{N}F_{j,i}\right)}$
    \EndFor
    \For{$i$ = $N+1$ to $N+M$}
    \State Move FC $i$ to
    $\frac{\sum_{j=1}^{N}\!F_{j,i}p_j}{\sum_{j=1}^{N}F_{j,i}}$
    \EndFor
    \State Run Bellman-Ford algorithm and update the normalized flow matrix i.e., $\bS=\R(\bP,\bW)$
    \State Calculate the flow matrix $\bF(\bW,\bS)$ and sensor power coefficients $g_i(\bP,\bS)$, $\forall i\in\mathcal{I}_{S}$
    \State Update the power diagrams $\Vor_{i}(\bP,\bS)$, $\forall i\in\mathcal{I}_{S}$
    \State Calculate the new cost function $D_{new}=D(\bP, \bW, \bS)$
\doWhile{$\frac{D_{old}-D_{new}}{D_{old}}\geq\epsilon$}
\end{algorithmic}
\end{algorithm}

\begin{theorem}
RL Algorithm is an iterative improvement algorithm, i.e., the cost function is non-increasing and the algorithm converges.
\label{convergence}
\end{theorem}
The proof of Theorem \ref{convergence} is provided in Appendix \ref{Appendix:theorem2}.

\section{Performance Evaluation}\label{sec:simulation}
In this section, we provide the experimental results for a WASN including 4 FCs and 40 sensors.
To make a fair comparison, we use the same target region and density function as in \cite{JGTCOM,SJH}, i.e., $\Omega=[0,10]^2$ and $f(\omega) = \frac{1}{\int_{\Omega}dA} = 0.01$.
Other parameters are set as follows: $\beta=1$, $\rho=0.1$, $\kappa=1$, $\epsilon=10^{-6}$.

Note that there is no existing work except our previous paper \cite{JGTCOM} considering both sensing uncertainty and energy consumption.
Bellman-Ford Algorithm \cite{Ford,Bellman} is the best routing algorithm to minimize the total energy consumption, but it does not take node deployment into account.
To compare with Bellman-Ford Algorithm, we apply random deployment and Lloyd Algorithm \cite{Lloyd} for the node deployment part.
Random deployment + Bellman-Ford (RBF) employs Bellman-Ford Algorithm on 100 random node deployments and selects the best one.
Similarly, Lloyd + Bellman-Ford (LBF) first applies Lloyd Algorithm to both FCs and Sensors to obtain a node deployment with small cost, and then employs Bellman-Ford Algorithm to reduce the average power.
Furthermore, we compare RL with CL \cite{JGTCOM} which combines two Lloyd-like algorithms to optimize the node deployment with one-hop communications.

\begin{figure}[!htb]
\setlength\abovecaptionskip{3pt}
\setlength\belowcaptionskip{0pt}
\centering
\includegraphics[width=3.3in]{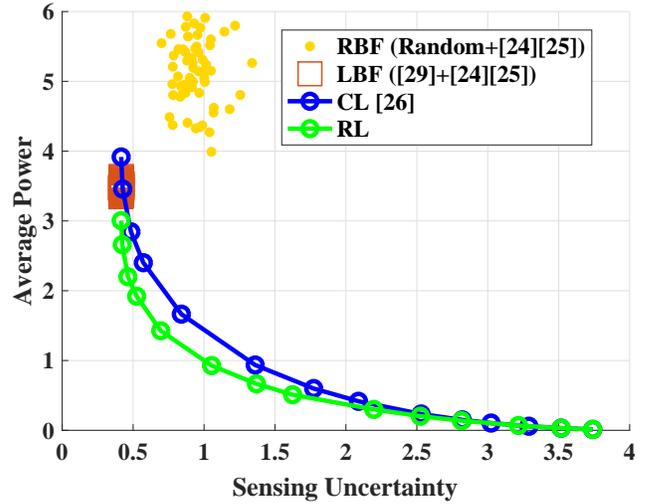}
\captionsetup{justification=justified}
\caption{\small{Performance comparison for RBF, LBF, CL and RL algorithms.}}
\label{Performance}
\end{figure}

Performance comparisons\footnote{To better exhibit the performance of LBF, CL, RL, we do not show the results of RBF with excessive powers ($\overline{\mathcal{P}}>6$) in Fig. \ref{Performance}.} for different values of $\lambda\in \left\{0, 0.05, 0.15, 0.25, 0.5, 1, 1.5, 2, 3, 4, 5, 7, 10, 16\right\}$ are provided in Fig. \ref{Performance}.
Note that CL and RL can adjust the node deployment in terms of the Lagrangian multiplier $\lambda$ while RBF and LBF are independent of $\lambda$. In particular, since LBF determines the node deployment by Lloyd Algorithm before employing Bellman-Ford Algorithm, LBF's performance is almost independent of the initial deployments, and its experimental results in Fig. \ref{Performance} converge to a point with small sensing uncertainty but large power consumption.
Overall, the proposed RL algorithm saves more energy or reduces more sensing uncertainty compared to other algorithms. It also provides a trade-off between the average power and sensing uncertainty.

\begin{figure}[!htb]
\setlength\abovecaptionskip{0pt}
\setlength\belowcaptionskip{0pt}
\centering
\subfloat[]{\includegraphics[width=1.69in]{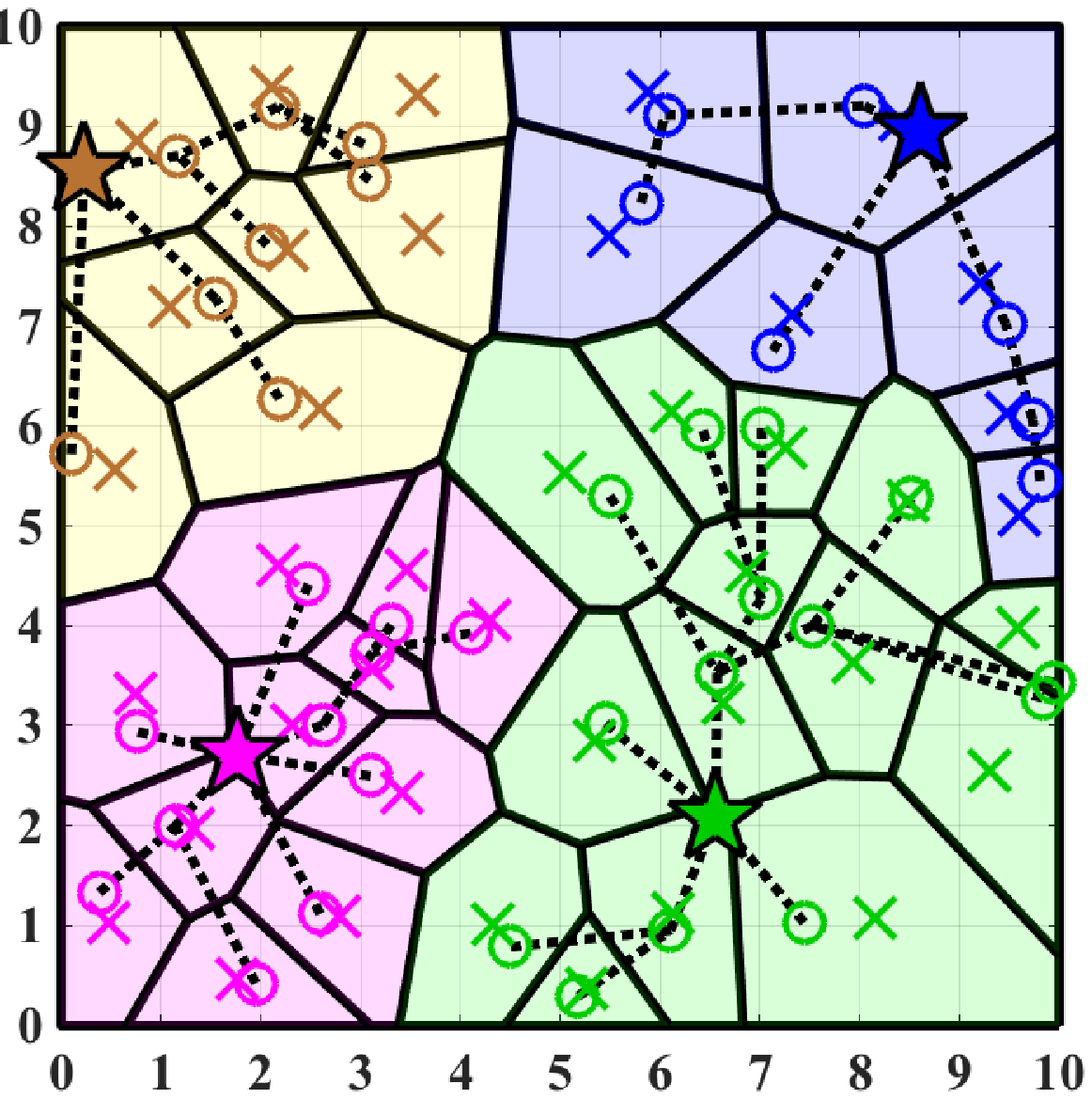}
\label{DeploymentWSN2a}}
\hfil
\subfloat[]{\includegraphics[width=1.69in]{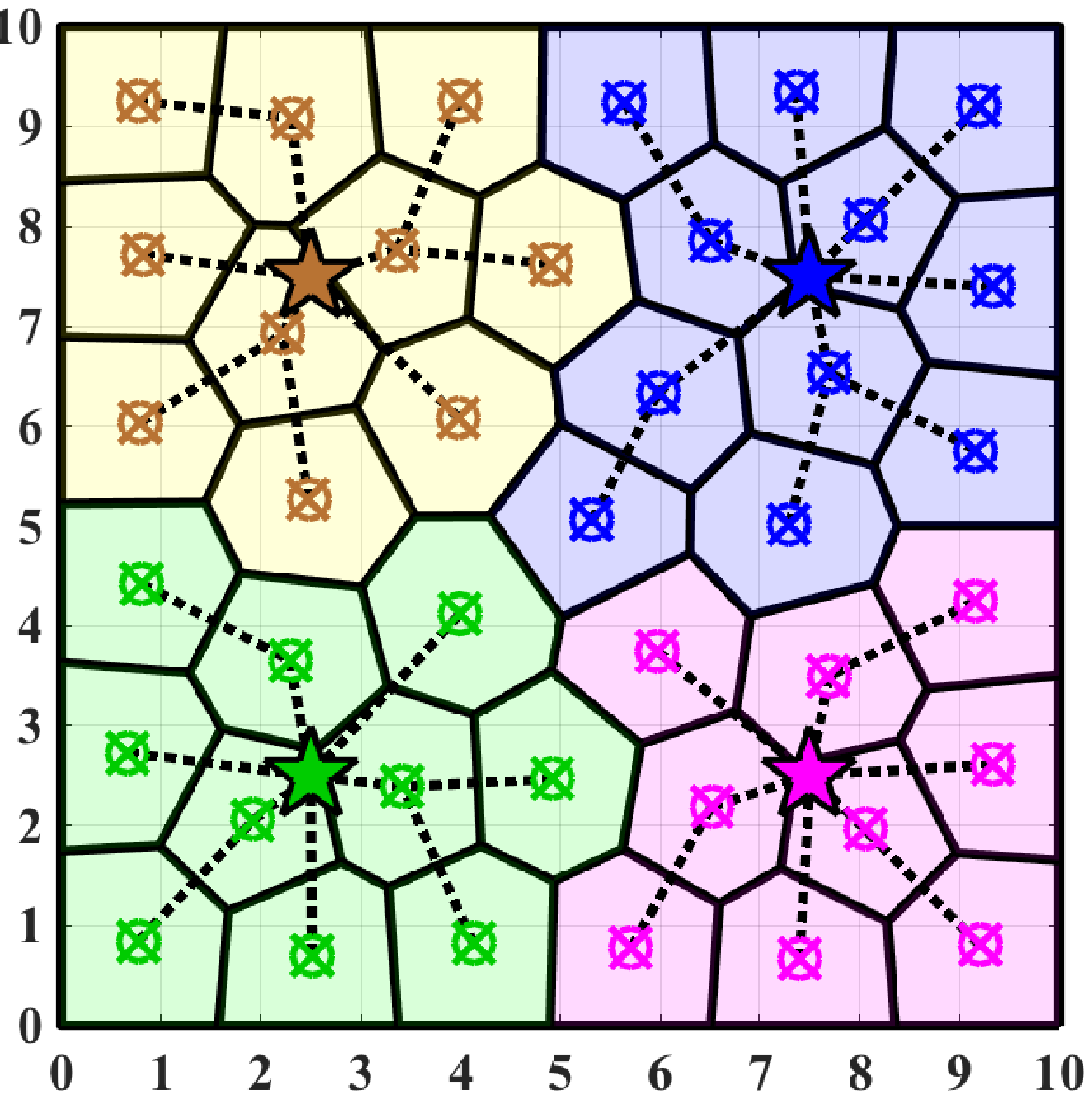}
\label{DeploymentWSN2b}}
\hfil
\subfloat[]{\includegraphics[width=1.69in]{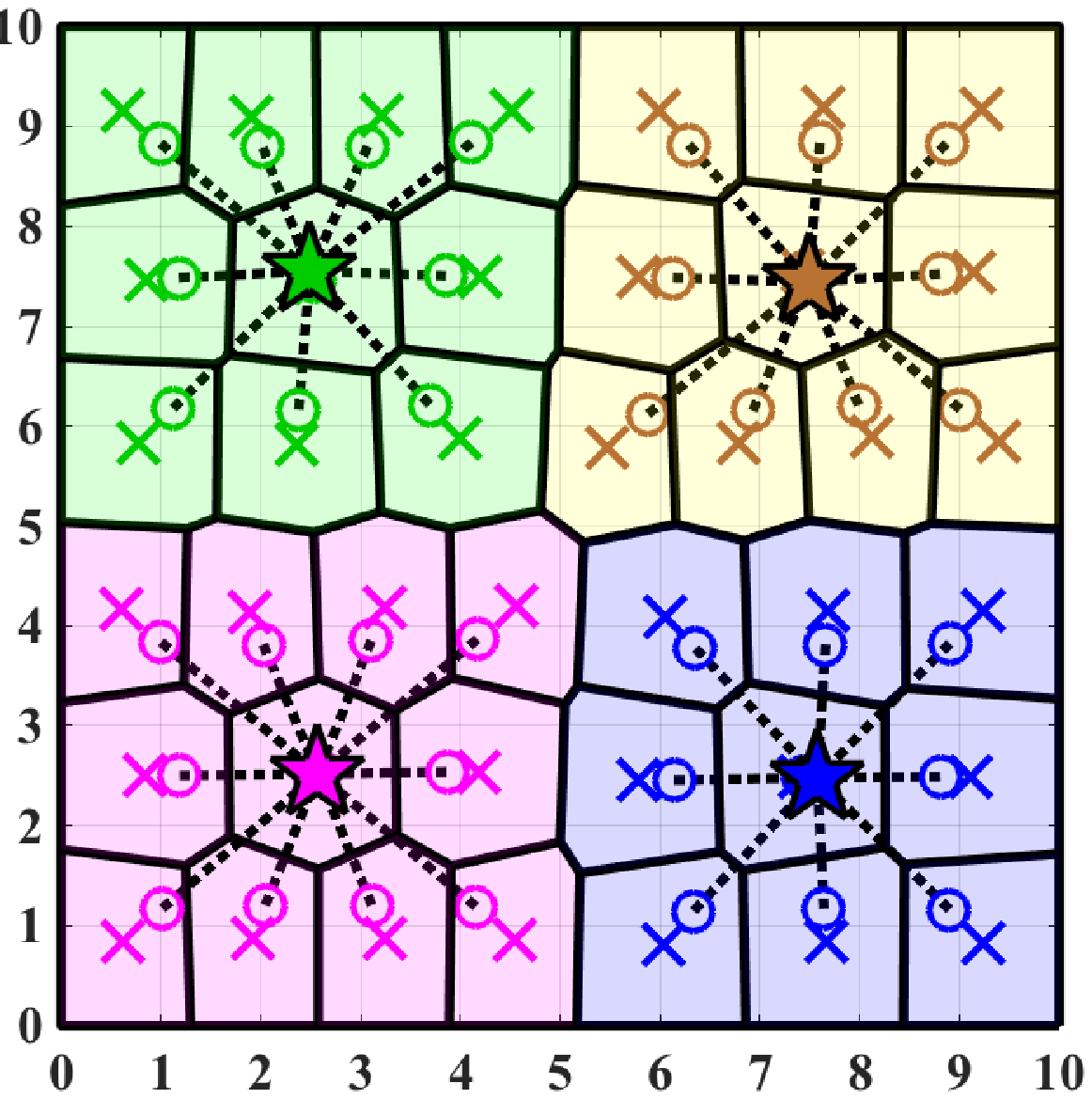}
\label{DeploymentWSN2c}}
\hfil
\subfloat[]{\includegraphics[width=1.69in]{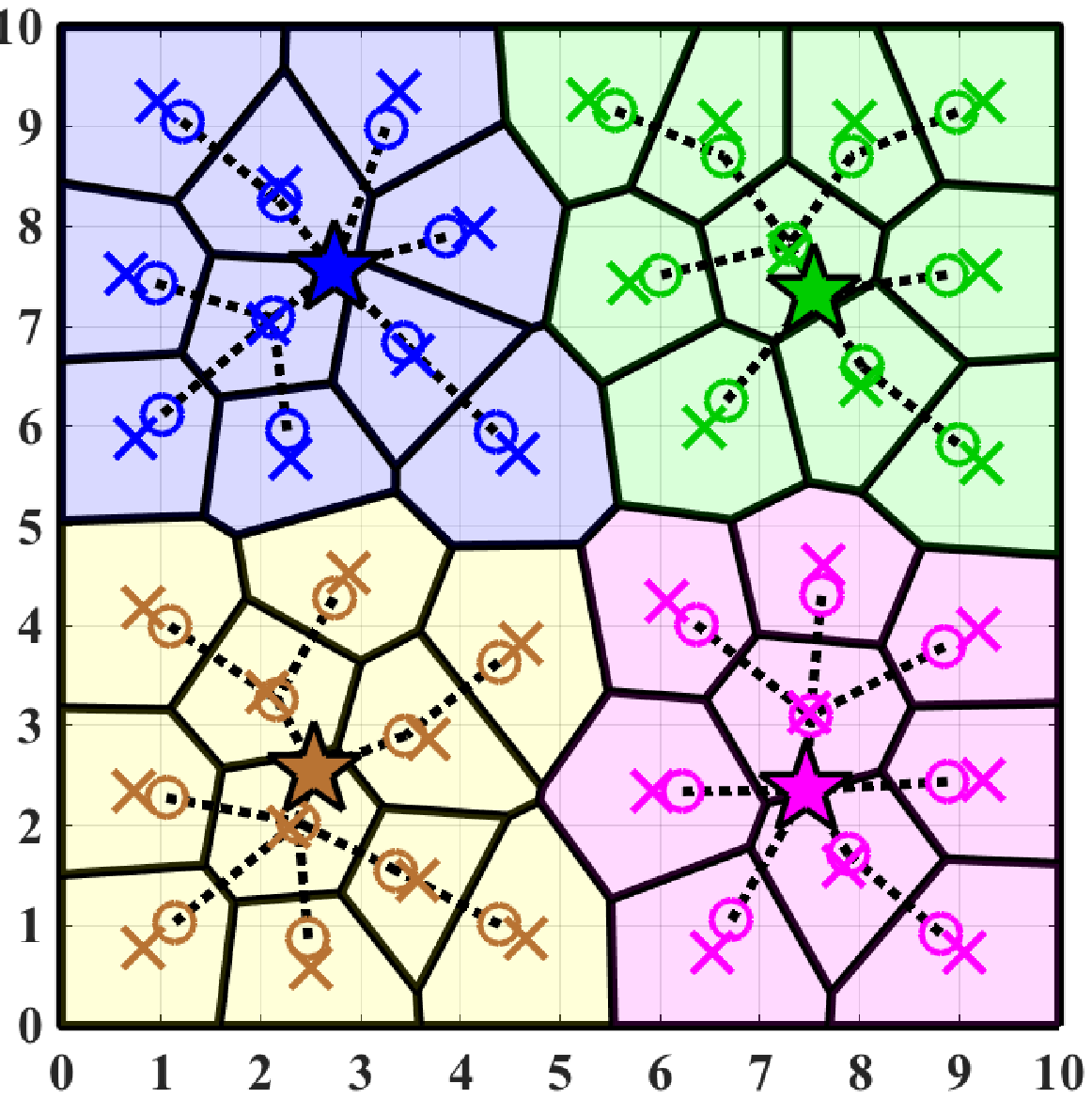}
\label{DeploymentWSN2d}}
\captionsetup{justification=justified}
\caption{\small{Node deployments of different algorithms with $\lambda=0.25$: (a) RBF (b) LBF (c) CL (d) RL.}}
\label{DeploymentInWSN2}
\end{figure}

The node deployments of the four algorithms (RBF, LBF, CL, and RL) in the WASN with $\lambda=0.25$ are illustrated in Figs. \ref{DeploymentWSN2a}, \ref{DeploymentWSN2b}, \ref{DeploymentWSN2c}, and \ref{DeploymentWSN2d}.
FCs and sensors are denoted by five-pointed stars and circles, respectively. Flows are denoted by black dotted lines. As shown in Fig. \ref{DeploymentInWSN2}, cell partitions in LBF, CL and RL algorithms tend to have similar shapes; however, RBF does not result in a similar pattern. Moreover, sensors in Fig. \ref{DeploymentWSN2b} are placed on top of their corresponding centroids while sensors in Fig. \ref{DeploymentWSN2c} are placed between their corresponding FC and centroid. However, in Fig. \ref{DeploymentWSN2d}, location of each sensor depends on its centroid, predecessors, and successors, as provided in Theorem 1. Note that in Figs. \ref{DeploymentWSN2b}, \ref{DeploymentWSN2c} and \ref{DeploymentWSN2d}, sensors inside each cluster tend to be close to each other with their FC in the middle; however, the same relationship does not appear in Fig. \ref{DeploymentWSN2a}. Besides, CL only uses one-hop communications, i.e., sensors are directly connected to the FC while other algorithms utilize multi-hop communications to reduce the average power.
The corresponding cost function given $\lambda=0.25$ for RBF, LBF, CL, and RL are, respectively, $1.87$, $1.25$, $1.17$, $1.01$; thus, our RL Algorithm achieves a lower cost function and outperforms other algorithms.

\section{Conclusions and Discussion}\label{sec:conclusion}
In this paper, we formulated the node deployment in WASNs as an optimization problem to make a trade-off between sensing uncertainty and energy consumption. The necessary conditions for the optimal deployment imply that each sensor location should be a linear combination of its centroid, predecessors and successors. Based on these necessary conditions, we proposed a Lloyd-like algorithm to minimize the total cost.
Our experimental results show that the proposed algorithm
significantly reduces both sensing uncertainty and energy consumption. Although we only considered Bellman-Ford Algorithm as the routing algorithm in this paper, the proposed system model in Section \ref{sec:model} can be applied to arbitrary routing algorithms, such as Flow Augmentation Algorithm \cite{JL} (a network lifetime maximization routing algorithm). The optimal deployment with maximum network lifetime will be our future work.

\appendices
\section{Proof of Theorem 1}\label{Appendix:theorem1}

Using parallel axis theorem, we can rewrite the cost function in (\ref{distortion1}) as:

\begin{equation}\label{parallel_axis_theorem}
\begin{aligned}
    D(\bP,\bW,\bS)=&\sum_{i=1}^{N}\int_{W_i}\|c_i-\omega\|^2f(\omega)d\omega+\|p_i-c_i\|^2v_i\\
    +&\lambda\rho\sum_{i=1}^{N}\sum_{j=1}^{N}F_{i,j}
    + \lambda\beta\sum_{i=1}^{N}\sum_{j=1}^{N+M}||p_i\!-\!p_j||^{2}F_{i,j},
\end{aligned}
\end{equation}

\noindent where $F_{i,j}=F_{i,j}(\bW,\bS)$, $v_i=\int_{W_i}f(\omega)d\omega$ is the Lebesgue measure (volume) of $W_i$ and $c_i=\frac{\int_{W_i}\omega f(\omega)d\omega}{\int_{W_i}f(\omega)d\omega}$ is the centroid of $W_i$. Let $\bP^*=(p^*_1,\dots,p^*_{N+M})^{T}$, $\bW^*=(W^*_1,\dots,W^*_{N})^{T}$, and $\bS^*=[s^*_{i,j}]$ denote, respectively, the optimal node deployment, cell partitioning and normalized flow matrix. According to \cite{VD}, each cell in the power diagram is either empty or a convex polygon; thus, we can take the gradient of the objective function $D\left(\bP,\bW,\bS \right)$ using Proposition A.1. in \cite{cortez}. It is self-evident that the cost function in (\ref{distortion1}) is continuously differentiable. Therefore, $D(\bP,\bW,\bS)$ achieves zero-gradient at the optimal point ($\bP^*$, $\bW^*$, $\bS^*$). The partial derivative of (\ref{parallel_axis_theorem}) with respect to $p_i$ is provided in (\ref{pd}), on top of the next page.
\begin{figure*}[!hbt]
\begin{equation}
    \frac{\partial D(\bP,\bW,\bS)}{\partial p_i}=
    \begin{cases}
    2(p_i-c_i)v_i +2\lambda\beta\sum\limits_{j=1}^{N+M}(p_i-p_j)F_{i,j}
    +2\lambda\beta\sum\limits_{j=1}^{N}(p_i-p_j)F_{j,i}, & \qquad \forall i\in \mathcal{I}_{S}\\
    2\lambda\beta\sum\limits_{j=1}^{N}(p_i-p_j)F_{j,i}, & \qquad \forall i\in \mathcal{I}_{F}
    \end{cases}
    \label{pd}
\end{equation}
\end{figure*}
By solving the zero-gradient equation, we obtain:

\begin{equation}
p^*_i=
\begin{cases}
\frac{c_i^*v_i^*+\lambda\beta\sum\limits_{j=1}^{N+M}F_{i,j}^*p^*_j+\lambda\beta\sum\limits_{j=1}^{N}F_{j,i}^*p^*_j}{v_i^*+\lambda\beta\left(\sum\limits_{j=1}^{N+M}F_{i,j}^*  +\sum\limits_{j=1}^{N}F_{j,i}^*\right)}, &\quad i\in\mathcal{I}_{S}\\
\frac{\sum\limits_{j=1}^{N}F_{j,i}^*p_j^*}{\sum\limits_{j=1}^{N}F_{j,i}^*},  &\quad i\in\mathcal{I}_{F}\\
\end{cases}
\label{zg}
\end{equation}
where $v_i^*$ and $c_i^*$ are, respectively, the volume and centroid of $W_i^*$, and $F_{i,j}^* = F_{i,j}\left(\bW^*,\bS^*  \right)$.

As shown at the beginning of Sec. \ref{sec:opt}, given the optimal deployment $\bP^*$ and the optimal normalized flow matrix $\bS^*$, the optimal cell partitioning is given by the power diagram $\bW^*=\Vor(\bP^*,\bS^*)$, indicating (\ref{optW}). Similarly, given the optimal deployment $\bP^*$ and the optimal cell partitioning $\bW^*$, the optimal normalized flow matrix is $\bS^*=\R(\bP^*,\bW^*)$, indicating (\ref{optS}).
Substituting (\ref{optW}) and (\ref{optS}) into (\ref{zg}), we get (\ref{optP}) and (\ref{optQ}) and the proof is complete.$\hfill\blacksquare$

\section{Proof of Theorem 2}
\label{Appendix:theorem2}

Note that when $\bW$, $\bS$, and $\{p_j\}_{j\ne i}$ are fixed, the cost function in (\ref{parallel_axis_theorem}) is a convex function of $p_i$; thus, by solving the zero-gradient equation, we have the following unique minimizer:
\begin{equation}
p_i=
\begin{cases}
\frac{c_iv_i+\lambda\beta\sum\limits_{j=1}^{N+M}F_{i,j}p_j+\lambda\beta\sum\limits_{j=1}^{N}F_{j,i}p_j}{v_i+\lambda\beta\left(\sum\limits_{j=1}^{N+M}F_{i,j}+\sum\limits_{j=1}^{N}F_{j,i}\right)}, & i\in\mathcal{I}_{S}\\
\frac{\sum\limits_{j=1}^{N}F_{j,i}p_j}{\sum\limits_{j=1}^{N}F_{j,i}}, & i\in\mathcal{I}_{F}
\end{cases}
\end{equation}
where $c_i$ and $v_i$ are centroid and volume of $W_i$, respectively. Therefore, moving sensors and FCs according to Lines 10 and 13 of Algorithm 1 does not increase the cost function. Since $\R(\bP,\bW)$ is the optimal normalized flow matrix for a given node deployment $\bP$ and cell partitioning $\bW$, updating $\bS$ according to Line 15 of Algorithm 1 does not increase the cost function either. As mentioned earlier, given the node deployment $\bP$ and normalized flow matrix $\bS$, the optimal cell partitioning is given by the power diagram $\Vor\left(\bP,\bS\right)$; hence, updating the cell partitioning according to Line 17 of Algorithm 1 also does not increase the cost function. Since the parameters $\bP$, $\bW$ and $\bS$ are updated only in Lines 10, 13, 15 and 17 of RL Algorithm, the cost function is non-increasing. In addition, the cost function is lower bounded by 0, i.e., $D\left(\bP,\bW,\bS \right)\geq0$. As a result, RL Algorithm is an iterative improvement algorithm and it converges.$\hfill\blacksquare$

\section*{Acknowledgment}
This work was supported in part by the NSF Award CCF-1815339.


\end{document}